\documentclass[11pt]{article}
\usepackage{amsmath,amsthm,amssymb}
\usepackage{graphicx}
\usepackage{array}
\usepackage[mathscr]{eucal}
\usepackage{eurosym}
\usepackage{xcolor}
\usepackage{amsmath}
\usepackage{booktabs} %% \addlinespace[1cm]  ; \toprule \midrule etc.

\usepackage{chngcntr}
\usepackage{siunitx} % S: align by decimal point (part of the siunitx package)

\usepackage{amsthm}
\theoremstyle{definition} % Non-italic type theorem
\newtheorem{algo}{Algorithm} % For Algorithm sections...
\newcommand{\mcc}[1]{\multicolumn{1}{c}{{#1}}}

\theoremstyle{definition}

\theoremstyle{remark}

\DeclareMathOperator{\Tr}{Tr} % Trace operator

\newcommand{\E}{\text{E}}

\newcommand{\V}{\text{Var}}

\newcommand{\VaR}{\text{VaR}}

\newcommand{\bi}{ \begin{itemize}  }
\newcommand{\ei}{\end{itemize}}

\newcommand{\bfor}{ \begin{eqnarray*} }
\newcommand{\efor}{\end{eqnarray*}}

\newcommand{\Mvec}{\mathbf{M}}

\newcommand{\Dvec}{\mathbf{D}}
\newcommand{\Cvec}{\mathbf{C}}
\newcommand{\Bvec}{\mathbf{B}}

\newcommand{\Pvec}{\mathbf{P}}

\newcommand{\Bmat}{\mathbf{B}}
\newcommand{\Dmat}{\mathbf{D}}
\newcommand{\Pmat}{\mathbf{P}}

\newcommand{\Wmat}{\mathbf{W}}

\newcommand{\kvec}{\mathbf{k}}

\newcommand{\mvec}{\mathbf{m}}

\newcommand{\gammavec}{\boldsymbol{\gamma}}

\newcommand{\thetavec}{\boldsymbol{\theta}}

\newcommand{\pivec}{\boldsymbol{\pi}}

\newcommand{\rit}{{\rm I\!R}}

\newcommand{\N}[2]{{\cal N}\left(#1,#2\right)}
\newcommand{\Nc}{{\cal N}}
\newcommand{\G}[2]{{\cal G }\left(#1,#2\right)}
\newcommand{\D}{{{\cal D}}}

\DeclareMathAlphabet\mathbfcal{OMS}{cmsy}{b}{n}

\newcommand{\argmax}{\arg\!\max}

\usepackage[latin1]{inputenc}
\usepackage[backend=biber,
    style=chicago-authordate,
    maxbibnames=5,
    sorting=nyt,
    url=true,
    natbib=true,
    hyperref=true,
    firstinits=true]{biblatex}

\begin{document}

\title{\bf Moment-based density and risk estimation \\
  from grouped summary statistics}

\author{{\Large P}HILIPPE {\Large L}AMBERT$^{1,2}$
  \\[5mm]
  $^1$ Institut de Recherche en Sciences Sociales (IRSS),\\
  M\'ethodes Quantitatives en Sciences Sociales,
  Universit\'e de Li\`ege, \\
  Place des Orateurs 3, B-4000 Li\`ege, Belgium\\
  Email: p.lambert@uliege.be
  \\[3mm]
$^2$ Institut de Statistique, Biostatistique et Sciences Actuarielles
(ISBA), \\ Universit\'e catholique de Louvain, \\
Voie du Roman Pays 20, B-1348 Louvain-la-Neuve, Belgium.\\
}
\maketitle

\begin{abstract}
  Data on a continuous variable are often summarized by means of
  histograms or displayed in tabular format: the range of data is
  partitioned into consecutive interval classes and the number of
  observations falling within each class is provided to the analyst.
  Computations can then be carried in a nonparametric way by assuming
  a uniform distribution of the variable within each partitioning
  class, by concentrating all the observed values in the center, or by
  spreading them to the extremities. Smoothing methods can also be
  applied to estimate the underlying density or a parametric model can
  be fitted to these grouped data.  For insurance loss data, some
  additional information is often provided about the observed values
  contained in each class, typically class-specific sample moments
  such as the mean, the variance or even the skewness and the
  kurtosis. The question is then how to include this additional
  information in the estimation procedure. The present paper proposes
  a method for performing density and quantile estimation based on
  such augmented information with an illustration on car insurance
  data.
  \\[5mm]
  \textit{Keywords:} Nonparametric density estimation, grouped data,
  sample moments, risk measures.
%\\[5mm]
%Subject classification: 60E15
\end{abstract}

%%%
\section{Introduction and motivation}

In risk analysis, 
losses are generally modelled as non-negative random variables and are
usually called {\em risks}. Analysts often need easy-to-compute
approximations of quantities relating to the risks they consider,
typically based on a few moments of the underlying loss
distribution. Several methods have been proposed in the literature,
including the classical Central Limit theorem, the Normal Power
approximation \parencite{Pentikainen:1987} based on Edgeworth
expansion or the maximum entropy principle \parencite{Brocketf:1995}
to cite a few.

Numerous moment bounds have also been developed in probability and
actuarial science.
Risk analysts indeed sometimes act in a conservative way by
basing their decisions on the least attractive risk that is consistent
with the incomplete available information (here, the range and the
first moments).  Since Markov fundamental inequality, a number of
improvements have been obtained under additional assumptions on the
underlying distribution function.  \textcite{BernardDenuitVanduffel:2014}
recently provided a new derivation of moment bounds on distribution
functions and Value-at-Risk measures, revisiting previous
contributions to the literature.  Besides distribution functions and
Value-at-Risk measures, bounds have also been derived on stop-loss
premiums and Tail-VaR, for instance.  \textcite{Hurlimann:2008} provides
a useful review of the available results.

In the present paper, we propose an efficient nonparametric estimation
procedure for the density based on histograms or grouped data,
including information about class-specific sample
moments. Specifically, the analyst has access to a set of data grouped
into consecutive classes (or tranches). Graphically, this corresponds
to an histogram. In addition to these grouped data, the average value
of the observations in each class is provided, as well as the
corresponding variance, skewness and kurtosis.

This format is often encountered in practice. For instance, in banking
and insurance contexts, operational risk loss data in the ORX annual
report (published by the operational risk management association) are
tranched and the total number of loss events as well as the total
gross loss falling within loss size boundaries are
provided. Reinsurers also often display the information about
insurance losses in this way. Confidentiality issues may sometimes
justify this grouping procedure. We show in this paper how to obtain a
smooth, nonparametric density estimate based on this information. It
is worth pointing out that the simulations conducted in the present
paper suggest that the additional information contained in
class-specific average values greatly improves the accuracy of the
estimation.  Of course, the proposed method can also be applied to
individual data.  It suffices to group them in an arbitrary number of
classes and to compute the corresponding sample moments.

The remainder of this paper is organized as follows. Section 2
formally describes the problem under investigation.  In Section 3, we
explain how to get a smooth estimate of the density based on summary
data.  As intermediate statistical goals, we also aim to quantify
uncertainty for the density estimate and derived quantities and to
evaluate the contribution of the different descriptive measures on the
density estimate (to issue recommendations for future reporting).
Section 4 is devoted to a simulation study assessing the performances
of the proposed approach.  In Section 5, we analyze a set of insurance
losses and we illustrate the value added of our new method.  The final
Section 6 discusses the results and research perspectives.

%%%
\section{Problem under investigation}

Our starting point is a set of $n$ observations $x_1,x_2,\ldots,x_n$
available in tabular form.  We consider that these observations are
realizations of independent random variables $X_1,X_2,\ldots,X_n$ with
common distribution function $F$ and density function $f$.  Precisely,
the available data points $x_i$ have been partitioned into consecutive
class intervals $\mathcal{C}_j$, $j=1,\ldots,J$, also called
{\it tranches}. These classes are defined by $\mathcal{C}_j=(a_{j-1},a_j]$
where the cut points $a_0,a_1,\ldots,a_J$ satisfy
$$
a_0<\min x_i<a_1<\ldots <a_{J-1}<\max x_i<a_J.
$$
In addition to the number $n_j$ of observations belonging to
$\mathcal{C}_j$, we also have summary statistics about observations in
each class. Specifically, we assume that we know the class-specific
means
$$
\overline{x}_j=\frac{1}{n_j}\sum_{x_i\in\mathcal{C}_j}x_i,\hspace{2mm}j=1,\ldots,J,
$$
as well as sample centered moments $m_{kj}$, $k=2,3,\ldots$, defined as
$$
m_{kj}=\frac{1}{n_j}\sum_{x_i\in\mathcal{C}_j}\big(x_i-\overline{x}_j\big)^k, \hspace{2mm}j=1,\ldots,J.
$$
Here, we consider the cases where variances, $s_j^2=m_{2j}$, skewness
coefficients, $g_{1j}=m_{3j}/m_{2j}^{3/2}$, and kurtosis,
$g_{2j}=m_{4j}/m_{2j}^{4/2}-3$, are available in addition to the means
$\overline{x}_j$.

Nonparametric computations are often carried out using the empirical
distribution function, assuming a uniform distribution of the class
relative frequencies $n_j/n$ over $\mathcal{C}_j$.  This allows the
risk analyst to estimate $\E [g(X)]$ by
$$
\sum_{j=1}^J\frac{n_j}{n(a_j-a_{j-1})}\int_{a_{j-1}}^{a_j}g(t)dt.
$$
This standard approach does not use any information about the
structure of the observed data inside each class.  Arbitrarily
assuming a uniform distribution of the losses in each risk class
$\mathcal{C}_j$ is contradicted by data for instance if
$\overline{x}_j\neq\frac{a_{j-1}+a_j}{2}$.

The approach proposed in the present paper integrates the information
about class-specific sample moments in the smooth density
estimate. Expectations of functions of $X$ are then easily computed,
as well as risk measures defined from quantiles such as Value-at-Risk,
\begin{align*}
&\VaR_\epsilon(X) = F^{-1}(1-\epsilon)=\inf\{x\in\mathbb{R}|F(x)\geq 1-\epsilon\},\hspace{2mm}\epsilon\in(0,1)
\end{align*}

%%%
\section{Methodology} \label{Methodology:Sec}

\subsection{Description of the model}  \label{ModelDescription:Sec}
Assume that each class $\mathcal{C}_j$ is divided into finer
sub-intervals.  The fine grid spacing $\Delta$ is taken small enough
to give an accurate description of the density $f$ for plotting it or
for computing quantiles or other indices accurately. The fine grid
consists in a sufficiently large number of grid points partitioning
$[a_0,a_J]$ into $I$ consecutive intervals
$\mathcal{I}_i=(b_{i-1},b_i]$ of equal width $\Delta$ with mid-point
$u_i$, $i=1,\ldots,I$.  For simplicity, assume that $\Delta$ is
selected in such a way that
$\{a_0,\ldots,a_J\}\subset \{b_0,\ldots,b_I\}$.  The relationship
between class $\mathcal{C}_j$ and the narrow bins
$\mathcal{I}_i$ is coded by means of the $J\times I$ matrix
$\Cvec = (c_{ji})$ where $c_{ji}=1$ if
$\mathcal{I}_i\subset\mathcal{C}_j$ and 0 otherwise.
\begin{figure}
   \centering
{\includegraphics[scale=0.6]{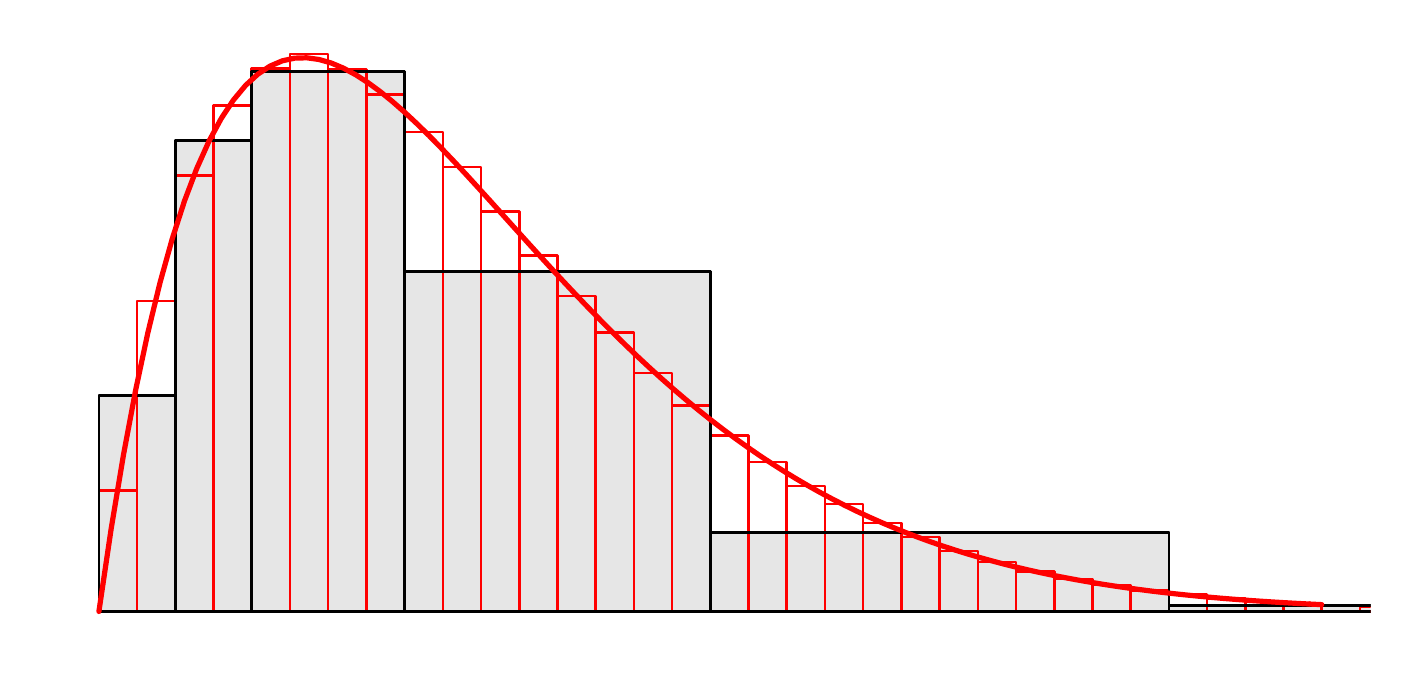}}
\caption{Density $f$ (continuous, red curve), latent distribution $\pivec$ and observed histogram (gray blocks).}
\label{FigfCD}
\end{figure}
Let
$$
\pi_i=\Pr(X\in\mathcal{I}_i)=\int_{b_{i-1}}^{b_i}f(t)dt = f(u_i)\Delta
+\mathcal{O}(\Delta^2),
$$
where $\pivec=(\pi_i)_{i=1}^I$ contains the values of the latent
distribution on the grid of $I$ narrow intervals $\mathcal{I}_i$
partitioning the support of $X$.  Figure \ref{FigfCD} illustrates the
construction (for a value of $\Delta$ much larger than what we use in
practice).  Consider a cubic B-spline basis $\{b_k(\cdot)\}_{k=1}^K$
associated to a large number of equidistant knots on $(a_0,a_J)$. We
model the probabilities in $\pivec$ using polytomous logistic
regression,
\begin{equation}
\label{RepPi}
\pi_i=\pi_i(\thetavec)=\frac{\exp(\eta_i)}{\exp(\eta_1)+\ldots+\exp(\eta_I)}
\end{equation}
where the scores $\eta_1,\ldots,\eta_I$ are connected to the B-spline
basis using
$$
\eta_i=\eta_i(\thetavec)=\sum_{k=1}^Kb_k(u_i)\theta_k=[\Bvec \thetavec]_i~,
$$
with each column of the matrix $[\Bvec]_{ik}=b_k(u_i)$ containing one
of the $K$ B-splines in the basis evaluated at the small bin
midpoints.  In this setting, we cannot observe $\pivec$ itself, but
only sums over $J<I$ intervals. The probability masses assigned to
these intervals are given by
$$
\Pr(X\in\mathcal{C}_j)=\int_{a_{j-1}}^{a_j}f(u)du=\gamma_j
,
$$
with $\gamma_j=\sum_{i=1}^Ic_{ji}\pi_i$, or in matrix form,
$\gammavec=\Cvec\pivec$, where $\Cvec$ is a $J\!\times I$ matrix. The
likelihood based on the observed grouped data frequencies,
$\D_0=\{n_1,\ldots,n_J\}$, directly follows from the Multinomial
distribution for the observed frequencies,
$\ell(\thetavec|\D_0)=\sum_{j=1}^J n_j \log\gamma_j$.
\textcite{EilersMarx:1996} proposed to put a discrete roughness penalty on
the B-splines coefficients to force smoothness on the density
estimate.  The penalized log-likelihood based on the observed data is
$$
\ell_{p}(\thetavec|\D_0,\lambda)
= \ell(\thetavec|\D_0) - \frac{\lambda}{2}\|\Dvec\thetavec\|^2,
$$
where $\Dvec$ is the $r$th order differencing matrix of size
$(K-r)\!\times K$ such that $\Dvec\thetavec =\Delta^r\thetavec$.  For
instance, with second-order differences, we have
$$
\Dmat=\begin{pmatrix}
1 & -2 & 1 & 0 & \cdots & 0 \\
0 & \ddots & \ddots & \ddots & \ddots & \vdots \\
\vdots & \ddots & \ddots & \ddots & \ddots & 0\\
0 & \cdots & 0 & 1 & -2 & 1 \\
\end{pmatrix}.
$$

\subsection{Estimation from grouped frequency data using the EM  algorithm}
\label{DensityEstimationFromFreqUsingEM:Sec}

The expectation-maximization (EM) algorithm
\parencite{DempsterLairdRubin:1977} is well-suited to estimate the
probabilities $\pivec$ from the latent (unobserved) small bin
frequencies $\mathbf{k}=(k_1,\ldots,k_I)$, where
$n_j=\sum_i c_{ji}k_i$.  Since $X_1,\ldots,X_n$ have been assumed to
be independent and identically distributed, the small bin frequencies
$k_1,\ldots,k_I$ are the realization of a Multinomial random vector
with exponent the sample size $n$ and probability vector $\pivec$. The
complete log-likelihood is then given by
$\ell^c(\thetavec|\D_0^c)=\sum_{i=1}^I k_i\log\pi_i(\thetavec)$ where
$\D_0^c=\{\mathbf{k}\}$. The penalized complete log-likelihood based
on the complete frequency data is
$$
\ell^c_{p}(\thetavec|\D_0^c,\lambda)
= \ell^c(\thetavec|\D_0^c) - \frac{\lambda}{2} \|\Dvec\thetavec\|^2.
$$
We propose to use Algorithm \ref{EM_FreqOnly:Algo} to perform estimation in the
described context.
%% Algorithm 1 ------>
\begin{algo} \label{EM_FreqOnly:Algo}
  {\it Density estimation using grouped data frequencies}\\
  The following EM algorithm alternates the update of the estimates
  for the latent frequencies $\mathbf{k}$, $\thetavec$ and, possibly,
  $\lambda$ using the following steps till convergence:
\begin{enumerate}
\item E-step: ${k}_i \leftarrow \E\left(k_i|\thetavec,\D_0\right)
=m_{\gamma(i)} \pi_i({\thetavec}) / \gamma_{j(i)}({\thetavec})$ where
$j(i)$ is such that ${\cal I}_i\subset \mathcal{C}_{j(i)}$ ;
\item M-step:
  ${\thetavec}=\argmax_\theta \ell^c_{p}(\thetavec|\D_0^c,\lambda)$.
  This can be done using penalized iteratively weighted least squares
  (P-IWLS) or a Newton-Raphson (N-R) algorithm. Let us detail the last
  algorithm.  Based on the following explicit forms for the gradient
  $\nabla_\theta {\ell}^c_{p}$ and the Hessian matrix
  ${\mathbf{H}_p^c}=\nabla^2_\theta {\ell}^c_{p}$,
  \begin{align}
    \nabla_\theta {\ell}^c_{p}(\thetavec|\D_0^c,\lambda) =
      \Bmat^\top(\kvec - n\pivec) - \lambda \Pmat\thetavec ~~;~
    -{\mathbf{H}_p^c} = -\nabla^2_\theta {\ell}^c_{p}(\thetavec|\D_0^c,\lambda)
                       =  \Bmat^\top\Wmat\Bmat + \lambda \Pmat~, \label{GradHes-FreqOnly:Eq}
  \end{align}
  where $\Pmat=\Dmat^\top\Dmat$ is the $K\!\times K$  penalty matrix and
  $(\Wmat)_{ii'}=n(\pi_{i}\delta_{ii'}-\pi_i\pi_{i'})$,
  the N-R algorithm repeats, till convergence, the following substitution:
  $
  \thetavec \longleftarrow \thetavec
  + \left(-{\mathbf{H}_p^c+\epsilon\,\mathrm{I}_K}\right)^{-1} \nabla_\theta {\ell}^c_{p}$.
  The addition of a small multiple of the identity matrix to the
  Hessian before inversion in the N-R step is a ridge penalty that
  conveniently handles the identification problem in (\ref{RepPi}),
  as $\pi_i(\thetavec)=\pi_i(\thetavec+c)$ for any constant $c$.
\item Penalty update:
 $\lambda \leftarrow ({\mathrm{edf}}-r)/\|\Dvec\thetavec\|^2$
where the effective number of spline parameters is given by the trace
of a matrix,
${\mathrm{edf}(\lambda)}
= \Tr\left( (-\mathbf{H}_p^c+\epsilon\,\mathrm{I}_K)^{-1}(-\mathbf{H}_p^c - \lambda\Pmat) \right)$.
\end{enumerate}
The last step is optional as one might prefer to perform the
estimation procedure for a given value of the penalty
parameter and rely on an external ad-hoc strategy to select
$\lambda$. At convergence, one obtains the penalized MLE
$\hat\thetavec_\lambda$ (given the selected value for the penalty
parameter $\lambda$). $\square$ %$\blacksquare$
\end{algo}
%% <------- End Algorithm 1
Estimation could also be performed using the strategy described in
\textcite{Eilers:2007} based on the composite link model
\parencite{ThompsonBaker:1981} or in a Bayesian setting, see Section
\ref{BayesianEstimation:Sec}.

\subsection{Estimation in a Bayesian setting} \label{BayesianEstimation:Sec}

\textcite{LambertEilers:2009} suggested to estimate $f$ using the
Bayesian paradigm.  In that context, the roughness penalty translates
into a smoothness prior for the spline coefficients,
$$
 p(\thetavec|\lambda)\propto \lambda^{\rho(\Pvec)/ 2}
 \exp\left(-{\lambda\over 2}~{\thetavec}^\top\Pvec{\thetavec}\right)
$$
where $\Pvec=\Dvec^\top\Dvec$ and $\rho(\Pvec)$ is the rank of
$\Pvec$.  A Gamma prior $\G{a}{b}$ with large variance for $\lambda$ is a
possible choice to express prior ignorance about suitable values for
$\lambda$, although more robust results can be obtained with a mixture
of Gammas
\parencite{JullionLambert:2007}.
Closed forms for the joint posterior of $(\thetavec,\lambda)$,
\begin{align}
&p(\thetavec,\lambda|\D_0)
\propto
\lambda^{a+\rho(\Pvec)/2-1}
\prod_j \gamma_j^{n_j} \exp\left\{-\lambda (b+ .5~{\thetavec}^\top\Pvec{\thetavec})\right\}
\label{JointPosterior:Eq1}
\end{align}
and its gradient are available. The Langevin-Hastings algorithm
\parencite{Roberts:1998} can be used to get a
random sample, $\{({\thetavec}^{(m)},\lambda^{(m)}), m=1,\ldots,M\}$,
from the posterior.  To each ${\thetavec}^{(m)}$ corresponds a density
$f^{(m)}$ from which any summary measure $\xi^{(m)}$ of interest such
as the mean, the standard deviation or quantiles can be computed.
Point estimates and credible intervals for $\xi$ can be derived from
$\{\xi^{(m)}$, $m=1,\ldots,M\}$.  Specific properties such as
unimodality or log-concavity can be imposed on the estimated density
by excluding, through the prior, the configurations of $\thetavec$
corresponding to non-desirable densities.  Alternatively, in that
Bayesian framework, Laplace approximations can be combined in a
Bayesian setting to estimate a density from grouped data in a fast and
reliable way, see \textcite[][Section 2.5] {Lambert:2021}. The
marginal posterior for the spline parameters is very well approximated
by a mixture of Normal distributions with weights defined by the
marginal posterior for the log of the penalty parameter. The latter
distribution can be reliably approximated by a skewed Normal
distribution \parencite{GressaniLambert:2021}.

One can show that maximizing (\ref{JointPosterior:Eq1}) for a given
value of $\lambda$ is equivalent to maximizing
$p(\thetavec|\lambda,\D_0)$ or the penalized complete log-likelihood
$\ell^c_{p}(\thetavec|\D_0^c,\lambda)$. Therefore, the so-obtained
conditional posterior mode coincides with the penalized MLE
$\hat\thetavec_\lambda$ given by the EM algorithm.
Further extensions to take into account the moments observed within
classes will be based on the EM algorithm.

\subsection{Moment-based extensions}
\label{DensityEstimationFromMomentsUsingEM:Sec}
\subsubsection{Density estimation for given class-specific sample means}

Let us now further assume that,  together with the frequencies $n_j$,
the sample means $\overline{x}_j$ within $\mathcal{C}_j$
($j=1,\ldots,J$) are also reported.
Then, the likelihood based on the observed data,
$\D_1=\D_0\cup \{\bar{x}_j:j=1,\ldots,J\}$,
becomes
$$
L(\thetavec|\D_1)
= \Pr\left(N_1=n_1,\ldots,N_J=n_J\right)\prod_{j=1}^J f_{\overline{X}_j}(\overline{x}_j| n_j)
 \propto \prod_{j=1}^J\Big(\gamma_j^{n_j}f_{\overline{X}_j}(\overline{x}_j| n_j)\Big),
$$
where $f_{\overline{X}_j}(\cdot| n_j)$ is the conditional density of
$\overline{X}_j$ given the class frequency $N_j=n_j$.  Provided that
the class frequency $n_j$ is not too small, the Central Limit theorem
provides a reasonable approximation to
$f_{\overline{X}_j}(\cdot| n_j)$. Formally, denote the class-specific
population moments as
$$
\mu_{1j} = {1\over \gamma_j} \int_{a_{j-1}}^{a_j} x f(x)~dx ~~\text{and}~~
\sigma^2_j = {1\over \gamma_j} \int_{a_{j-1}}^{a_j} (x-\mu_{1j})^2 f(x) dx.
$$
Then,
$$
f_{\overline{X}_j}(\overline{x}_j| n_j)\approx
\frac{\sqrt{n_j}}{\sigma_j\sqrt{2\pi}}\exp\left(-\frac{n_j}{2\sigma_j^2}(\overline{x}_j-\mu_{1j})^2\right).
$$
Given the preceding spline approximation to the (log-)density using
polytomous logistic regression for the probabilities to be in the
small bins partitioning the support, see \eqref{RepPi}, one has
\begin{eqnarray*}
\mu_{1j}(\thetavec) & = & \E[X|X \in \mathcal{C}_j]
  = \sum_{i=1}^I u_i \frac{c_{ji}\pi_i(\thetavec)}{ \gamma_j(\thetavec)} +\mathcal{O}(\Delta^2), \\
\sigma^2_j(\thetavec) & = & \V[X|X \in\mathcal{C}_j]
  = \sum_{i=1}^I \big(u_i-\mu_{1j}(\thetavec)\big)^2~\frac{c_{ji}\pi_i(\thetavec)}{ \gamma_j(\thetavec)} +\mathcal{O}(\Delta^2).
\end{eqnarray*}
Then, for given roughness penalty parameter $\lambda$, the penalized
log-likelihood based on $\D_1$ becomes
$$
\ell_p(\thetavec|\D_1,\lambda) = \ell_p(\thetavec|\D_0,\lambda)
-{1\over 2} \sum_{j=1}^J \left(
\ln\sigma^2_j(\thetavec) + {n_j \over
  \sigma^2_j({\thetavec})} \big(\overline{x}_j-\mu_{1j}(\thetavec)\big)^2
\right),
$$
while its counterpart based on latent small bins frequencies,
$\D_1^c=\D_0^c\cup \{\bar{x}_j:j=1,\ldots,J\}$, is
\begin{align}
\ell_p^c(\thetavec|\D_1^c,\lambda) = \ell_p^c(\thetavec|\D_0^c,\lambda)
-{1\over 2} \sum_{j=1}^J \left(
\ln\sigma^2_j(\thetavec) + {n_j \over \sigma^2_j({\thetavec})} \big(\overline{x}_j-\mu_{1j}(\thetavec)\big)^2
\right). \label{PenLlik-MeanFreq:Eq}
\end{align}
It leads to Algorithm \ref{EM_FreqMean:Algo} for an estimation of the
density using the EM algorithm.

%% Algorithm 2 ------>
\begin{algo} \label{EM_FreqMean:Algo}
{\it Density estimation using grouped data means and frequencies}\\
The following EM algorithm alternates the update of the estimates
for the latent frequencies $\mathbf{k}$, $\thetavec$ and, possibly,
$\lambda$.
Denote by $\tilde\Bmat$ the $I\!\times K$ matrix such that
$(\tilde\Bmat)_{ik}= \tilde{b}_{ik}
= b_k(u_i)-\sum_{t=1}^I b_k(u_{t})\pi_t$.

\vspace{.2cm} \noindent
Repeat the following steps till convergence:
\begin{enumerate}
\item E-step: ${k}_i \leftarrow \E\left(k_i|\thetavec,\D_1\right)
=m_{\gamma(i)} \pi_i({\thetavec}) / \gamma_{j(i)}({\thetavec})$ where
$j(i)$ is such that ${\cal I}_i\subset \mathcal{C}_{j(i)}$ ;
\item M-step: compute
  ${\thetavec}=\argmax_\theta
  \ell^c_{p}(\thetavec|\D_1^c,\lambda)$.
  This can be done using a Newton-Raphson (N-R) algorithm with
  the iterative substitution,
  $
  \thetavec \longleftarrow \thetavec
  + \left({-\mathbf{H}_p^c+\epsilon\,\mathrm{I}_K}\right)^{-1} \nabla_\theta {\ell}^c_{p}
  $, till convergence. Explicit forms for the gradient
  and the Hessian matrix
  are available after conditioning on the value
  $\tilde\sigma^2_j$ of $\sigma^2_j(\thetavec)$ in (\ref{PenLlik-MeanFreq:Eq}) at the start of the iteration:
  \begin{align}
    \begin{split}
    &\nabla_\theta {\ell}^c_{p}(\thetavec|\D_1^c,\lambda) =
    \nabla_\theta {\ell}^c_{p}(\thetavec|\D_0^c,\lambda)
    + \sum_{j=1}^J {n_j \over \tilde\sigma^2_j}
       \big(\overline{x}_j-\mu_{1j}\big) {\partial \mu_{1j}\over \partial\thetavec}
      ~~;~ \\
    &-{\mathbf{H}_p^c} = -\nabla^2_\theta {\ell}^c_{p}(\thetavec|\D_1^c,\lambda)
      \approx -\nabla^2_\theta {\ell}^c_{p}(\thetavec|\D_0^c,\lambda)
    + \sum_{j=1}^J {n_j \over \tilde\sigma^2_j}
       {\partial \mu_{1j}\over \partial\thetavec} {\partial \mu_{1j}\over \partial\thetavec^\top},
     \end{split}
    \label{GradHes-MeanFreq:Eq}
  \end{align}
  where
  $
  {\partial \mu_{1j} / \partial\theta_k}
  ={1 \over \gamma_j}\sum_{i=1}^I c_{ji}\pi_i(u_i-\mu_{1j})b_{ik}
  $,
  with the approximation to $\mathbf{H}_p^c$ coming from the neglect
  of zero expectation terms.

\item Penalty update:
 $\lambda \leftarrow ({\mathrm{edf}}-r)/\|\Dvec\thetavec\|^2$
where the effective number of spline parameters is given by the trace
of a matrix,
${\mathrm{edf}(\lambda)}= \Tr\left(
                  (-\mathbf{H}_p^c+\epsilon\,\mathrm{I}_K)^{-1}(-\mathbf{H}_p^c - \lambda\Pmat)
                 \right)$.
\end{enumerate}
At convergence, one obtains the penalized MLE
$\hat\thetavec_\lambda$ (given the selected value for $\lambda$). $\square$ %$\blacksquare$
\end{algo}
%% <------- End Algorithm 2

%%
\subsubsection{Density estimation for given class-specific sample central moments}

Assume now that, together with the frequencies,
the sample mean $\overline{X}_j$, variance $S^2_j$, skewness $G_{1j}$
and kurtosis $G_{2j}$ within $\mathcal{C}_j$
 are reported. If $M_{rj}$ denotes the $r$th sample central moment in $\mathcal{C}_j$, we have
$$
\overline{X}_j=M_{1j}~~;~~
S^2_j=M_{2j}~~;~~
G_{1j}= M_{3j} / {M_{2j}^{3/2}}~~;~~
G_{2j}= M_{4j} / {M_{2j}^{4/2}}-3.
$$
Denote by $\Mvec_j=( M_{1j} , M_{2j} , M_{3j}, M_{4j})$ the random vector of
sample central moments in $\mathcal{C}_j$ and by $\mvec_j$ its
observed counterpart. For sufficiently large values of $n_j$, the
Central Limit theorem provides a multivariate Normal approximation to
the sampling distribution of $\Mvec_j$,
$\Mvec_j \overset{d}{\longrightarrow}
\mathcal{N}_4\left(\pmb{\mu}_j,\Sigma_j\right),$
where $\pmb{\mu}_j=(\mu_{1j},\ldots,\mu_{4j})$ with $\mu_{1j}=\mu_j$ and
\begin{eqnarray*}
\mu_{rj} = \mu_{rj}(\thetavec) &=& \E\left[
(X-\mu_{1j}(\thetavec) )^r | X \in \mathcal{C}_j\right],\hspace{2mm}r=2,3,\ldots\\
&=& {1\over \gamma_j(\thetavec)} \sum_{j=1}^J c_{ji}
           \pi_i(\thetavec) \big(u_i-\mu_{1j}(\thetavec)\big)^r +\mathcal{O}(\Delta^2).
\end{eqnarray*}
Using the Generalized Method of Moments (GMM) \parencite{Hansen:1982}, one can show that
\begin{align}
\Sigma_j =
{1\over n_j}
\begin{pmatrix}
\mu_{2j} & \mu_{3j} & \mu_{4j} &  \mu_{5j} \\
\mu_{3j} & \mu_{4j}-\mu_{2j}^2 & \mu_{5j}-\mu_{2j}\mu_{3j} & \mu_{6j}-\mu_{2j}\mu_{4j} \\
\mu_{4j} & \mu_{5j}-\mu_{2j}\mu_{3j} & \mu_{6j}-\mu_{3j}^2 & \mu_{7j}-\mu_{3j}\mu_{4j} \\
\mu_{5j} & \mu_{6j}-\mu_{2j}\mu_{4j} & \mu_{7j}-\mu_{3j}\mu_{4j} & \mu_{8j}-\mu_{4j}^2
\end{pmatrix}.  \label{Sigma_j:Eq}
\end{align}
Based on the observed data $\D=\D_0\cup \{\mvec_j:j=1,\ldots,J\}$ and
for a given roughness penalty parameter $\lambda$, the penalized log-likelihood
becomes
\begin{align}
\ell_p(\thetavec|\D,\lambda) = \ell_p(\thetavec|\D_0,\lambda)
-{1\over 2} \sum_{j=1}^J \left\{
 \ln |\Sigma_j|
+(\mvec_j-\pmb{\mu}_j)^\top \Sigma_j^{-1} (\mvec_j-\pmb{\mu}_j)
\right\},  \label{PenLlik-FreqCentralMoments:Eq}
\end{align}
comparable to (\ref{PenLlik-MeanFreq:Eq}) when only the tabulated
sample means were available. When, in addition, the latent small bins
frequencies are given, $\D^c=\D^c_0\cup \{\mvec_j:j=1,\ldots,J\}$,
inference is based on the penalized complete log-likelihood,
\begin{align}
\ell^c_p(\thetavec|\D^c,\lambda) = \ell^c_p(\thetavec|\D^c_0,\lambda)
-{1\over 2} \sum_{j=1}^J \left\{
 \ln |\Sigma_j|
+(\mvec_j-\pmb{\mu}_j)^\top \Sigma_j^{-1} (\mvec_j-\pmb{\mu}_j)
\right\}. \label{CompletePenLlik-FreqCentralMoments:Eq}
\end{align}
The maximization of the penalized log-likelihood
(\ref{PenLlik-FreqCentralMoments:Eq}) and the selection of $\lambda$
can be made using Algorithm \ref{EM_FreqCentralMoments:Algo}.

%% Algorithm 3 ------>
\begin{algo} \label{EM_FreqCentralMoments:Algo}
{\it Density estimation given class-specific sample moments and frequencies}\\
The following EM algorithm alternates the update of the estimates
for the latent frequencies $\mathbf{k}$, $\thetavec$ and, possibly, $\lambda$ using the
following steps till convergence:
\begin{enumerate}
\item {E-step}: ${k}_i \leftarrow \E\left(k_i|\thetavec,\D_1\right)
=m_{\gamma(i)} \pi_i({\thetavec}) / \gamma_{j(i)}({\thetavec})$ where
$j(i)$ is such that ${\cal I}_i\subset \mathcal{C}_{j(i)}$ ;
\item M-step: compute
  ${\thetavec}=\argmax_\theta
  \ell^c_{p}(\thetavec|\D^c,\lambda)$.
  This can be done using a Newton-Raphson (N-R) algorithm with
  the iterative substitution,
  $
  \thetavec \longleftarrow \thetavec
  + \left({-\mathbf{H}_p^c+\epsilon\,\mathrm{I}_K}\right)^{-1} \nabla_\theta {\ell}^c_{p}
  $, till convergence. Explicit forms for the gradient
  and the Hessian matrix
  are available after conditioning on the value
  $\tilde\Sigma_j$ of $\Sigma_j(\thetavec)$ in
  (\ref{CompletePenLlik-FreqCentralMoments:Eq}) at the start of the
  iteration:
  \begin{align}
    \begin{split}
    \big(\nabla_\theta {\ell}^c_{p}\big)_k &=
    \big(\nabla_\theta {\ell}^c_{p}(\thetavec|\D_0^c,\lambda)\big)_k
    + \sum_{j=1}^J (\mvec_j-\pmb{\mu}_j)^\top \tilde\Sigma_j^{-1}
        {\partial \pmb{\mu}_j\over \partial\theta_k}
      ~~;~ \\
    -\big({\mathbf{H}_p^c}\big)_{ks} &\approx -\big(\nabla^2_\theta {\ell}^c_{p}(\thetavec|\D_0^c,\lambda)\big)_{ks}
    + \sum_{j=1}^J {\partial \pmb{\mu}^\top_j\over \partial\theta_s}
        \tilde\Sigma_j^{-1} {\partial \pmb{\mu}_j\over \partial\theta_k}
     \end{split}
    \label{GradHes-MeanFreq:Eq}
  \end{align}
  with
\begin{align}
  {\partial \mu_{rj} \over \partial\theta_k}
&  ={1 \over \gamma_j} \sum_{i=1}^I c_{ji}\pi_i b_{ik}
\big\{ (u_i-\mu_{1j})^r -\mu_{rj} \big\}
    -r \mu_{r-1,j}{\partial\mu_{1j}\over \partial\theta_k} \nonumber
  \\
&  ={1 \over \gamma_j} \sum_{i=1}^I c_{ji}\pi_i b_{ik} \big\{
    (u_i-\mu_{1j})^r -\mu_{rj}
    -r \mu_{r-1,j}(u_i-\mu_{1j})
 \big\} \label{DerivativeMoments:Eq}
\end{align}
for $1\leq k,s\leq K$ and $2\leq r\leq 4$.

\item Penalty update:
 $\lambda \leftarrow ({\mathrm{edf}}-r)/\|\Dvec\thetavec\|^2$
where the effective number of spline parameters is given by the trace
of a matrix,
${\mathrm{edf}(\lambda)}= \Tr\left(
                  (-\mathbf{H}_p^c+\epsilon\,\mathrm{I}_K)^{-1}(-\mathbf{H}_p^c - \lambda\Pmat)
                 \right)$.
\end{enumerate}
At convergence, one obtains the penalized MLE
$\hat\thetavec_\lambda$ (given the selected value for $\lambda$). $\square$ %$\blacksquare$
\end{algo}
%% <------- End Algorithm 3

\subsection{Quantile estimation} \label{QuantileEstimation:Sec}
Consider the following shorthand notation for the conditional
posterior mode of the vector of spline parameters underlying the
density estimate, ${\hat{\thetavec}}={\hat{\thetavec}}_\lambda - \max\{\hat\theta_k\}_{k=1}^K$,
with a substraction of the largest estimated component to handle the identification issue following from
$\pivec(\thetavec+c)=\pivec(\thetavec)$ for any real number $c$.
Denote by $\hat{k}$ the component for which $(\hat\thetavec)_{\hat{k}}=0$
and by $\thetavec_{-\hat{k}}$ the vector of spline parameters where the $\hat{k}$th component
of $\thetavec$ is  omitted. 
Quantile estimates can be derived using the fitted density estimate,
$$
f(x|\hat{\thetavec}) = {\exp\{\eta(x|\hat{\thetavec})\} /
   \int_{\cal X} \exp\{\eta(t|\hat{\thetavec})\}\,dt}\,,
$$
where
$\eta(x|{\thetavec}) = \sum_{k=1}^Kb_k(x)\theta_k$ and
${\cal X}=(a_0,a_J)$ denotes the support of the density.
Indeed, as the associated estimate for the cumulative distribution function (CDF),
$\hat{F}(x)=F(x|\hat{\thetavec}) = \int_{a_0}^x f(t|\hat{\thetavec})\,dt$,
is monotone, it can be inverted to provide an estimate of the quantile function,
$$
\hat{Q}(p) = Q(p|\hat{\thetavec}) = \inf\{x\in\rit: F(x|\hat{\thetavec})\geq p\}.
$$
Practically, starting for the fitted probability,
$\hat\pi_i=\Pr(X\in\mathcal{I}_i|\hat{\thetavec})$, to have an observation in
the small bin $\mathcal{I}_i=(b_{i-1},b_i]$ ($i=1,\ldots,I$), see Section \ref{ModelDescription:Sec},
and with $\hat{F}(b_0)=0$, $\hat{F}(b_i)=\hat\pi_1+\ldots+\hat\pi_i$, 
a first guess for $\hat{Q}(p)$ is given by
$$x_0=\max_{0\leq i\leq I}\big\{b_i: \hat{F}(b_i)\leq p\big\}\,.$$
This first approximation can be improved in an iterative way with, at iteration $t$,
$$x_t \longleftarrow x_{t-1} + \big(p - F(x_{t-1}|\hat{\thetavec})\big) / f(x_{t-1}|\hat{\thetavec})\,,$$
yielding at convergence $Q(p|\hat{\thetavec})=x_\infty$. The
uncertainty in that estimation directly follows from the uncertainty
in the choice of $\thetavec$. The latter is quantified by the conditional posterior distribution of 
${\thetavec}_{-\hat{k}}$ with Laplace approximation
$\big({\thetavec}_{-\hat{k}}|\D,\lambda\big) \stackrel{\cdot}{\sim}
\Nc\big({\hat{{\thetavec}}}_{-\hat{k}}, ({\cal J}_{-\hat{k},-\hat{k}})^{-1}\big)$
where
\begin{align}
  {\cal J}_{ks} &=
-\big(\nabla^2_{{\theta}} {\ell}_{p}(\thetavec|\D_0,\lambda)\big)_{ks}
+ \sum_{j=1}^J {\partial \pmb{\mu}^\top_j\over \partial\theta_s}
                   \tilde\Sigma_j^{-1} {\partial \pmb{\mu}_j\over \partial\theta_k}\,,
                   \label{ThetaMat:Eq}
\end{align}
with the partial derivatives of the theoretical central moments
$\pmb{\mu}_j$ in the $j$th class given in
(\ref{DerivativeMoments:Eq}).
The second derivatives of the log-likelihood based on the observed class frequencies $\D_0$
in the first term of (\ref{ThetaMat:Eq}) has an explicit form given by
\begin{align*}
  -\big(\nabla^2_\theta {\ell}_{p}(\thetavec|\D_0,\lambda)\big)_{ks}
  & = n \sum_{i=1}^I b_{ik}\pi_i\tilde{b}_{is} %\\
%  & ~~~
    + \sum_{j=1}^J {n_j\over\gamma_j²}\sum_{\ell=1}^I c_{j\ell}\pi_\ell \tilde{b}_{\ell s}
    \sum_{i=1}^I c_{ji}\pi_i \tilde{b}_{ik} %\\
%  & ~~~
     - \sum_{j=1}^J {n_j\over\gamma_j}\sum_{i=1}^I c_{ji}\pi_i \tilde{b}_{is}\tilde{b}_{ik} \\
  & = (\Bmat^\top\Wmat\Bmat)_{ks} % = n \sum_{i=1}^I b_{ik}\pi_i\tilde{b}_{is}
    - \sum_{j=1}^J {n_j\over\gamma_j}\sum_{i=1}^I c_{ji}\pi_i {b}_{ik}
      \left(b_{is}-{1\over\gamma_j}\sum_{\ell=1}^I c_{j\ell}\pi_\ell b_{\ell s}\right).
\end{align*}
The first term in the last expression, where
$(\Wmat)_{ii'}=n(\pi_{i}\delta_{ii'}-\pi_i\pi_{i'})$, corresponds to the information
available on $\thetavec$ based on data frequencies in the absence of class
tabulation. The information reduction due to tabulation is
quantified by the second term.
Based on the following first-order expansion,
$$Q(p|\thetavec) \approx Q(p|\hat\thetavec)
+ {\partial Q(p|\hat\thetavec) \over \partial \thetavec_{-\hat{k}}^\top}(\thetavec-\hat\thetavec)_{-\hat{k}}\,,
$$
the conditional posterior distribution of $Q(p|\thetavec)$ can be approximated by
\begin{align}
\big(Q(p|\thetavec)|\D,\lambda\big) \stackrel{\cdot}{\sim}\N{Q(p|\hat\thetavec)}
{s^2_Q(p)}\,,
  \label{QuantilePosterior:Eq}
\end{align}
with
\begin{align*}
{\partial Q(p|\hat\thetavec) \over \partial \theta_k} &=
{1\over f\big(Q(p|\hat\thetavec)|\hat\thetavec\big)}
\left\{
  \int_{-\infty}^{Q(p|\hat\thetavec)}b_k(x)f(x|\hat\thetavec)\,dx - p\int_\rit b_k(x)f(x|\hat\thetavec)\,dx
\right\}\,, \\
s^2_Q(p ) &= {\partial Q(p|\hat\thetavec) \over \partial \thetavec_{-\hat{k}}^\top}
\,({\cal J}_{-\hat{k},-\hat{k}})^{-1}\,
  {\partial Q(p|\hat\thetavec) \over \partial \thetavec_{-\hat{k}}}\,.
\end{align*}
Therefore, an approximate $100(1-\alpha)\%$ credible interval for $Q(p)$ is given by
\begin{align}
  & Q(p|\hat\thetavec) \pm \Phi^{-1}(1-\alpha/2)\,s_Q(p)\,,
      \label{QuantileCredibleInterval:Eq}
\end{align}
with $\Phi^{-1}(\cdot)$ denoting the quantile function of the standard Normal distribution.
        
\section{Simulation study}

\begin{figure}
   \centering
\includegraphics[width=.8\textwidth]{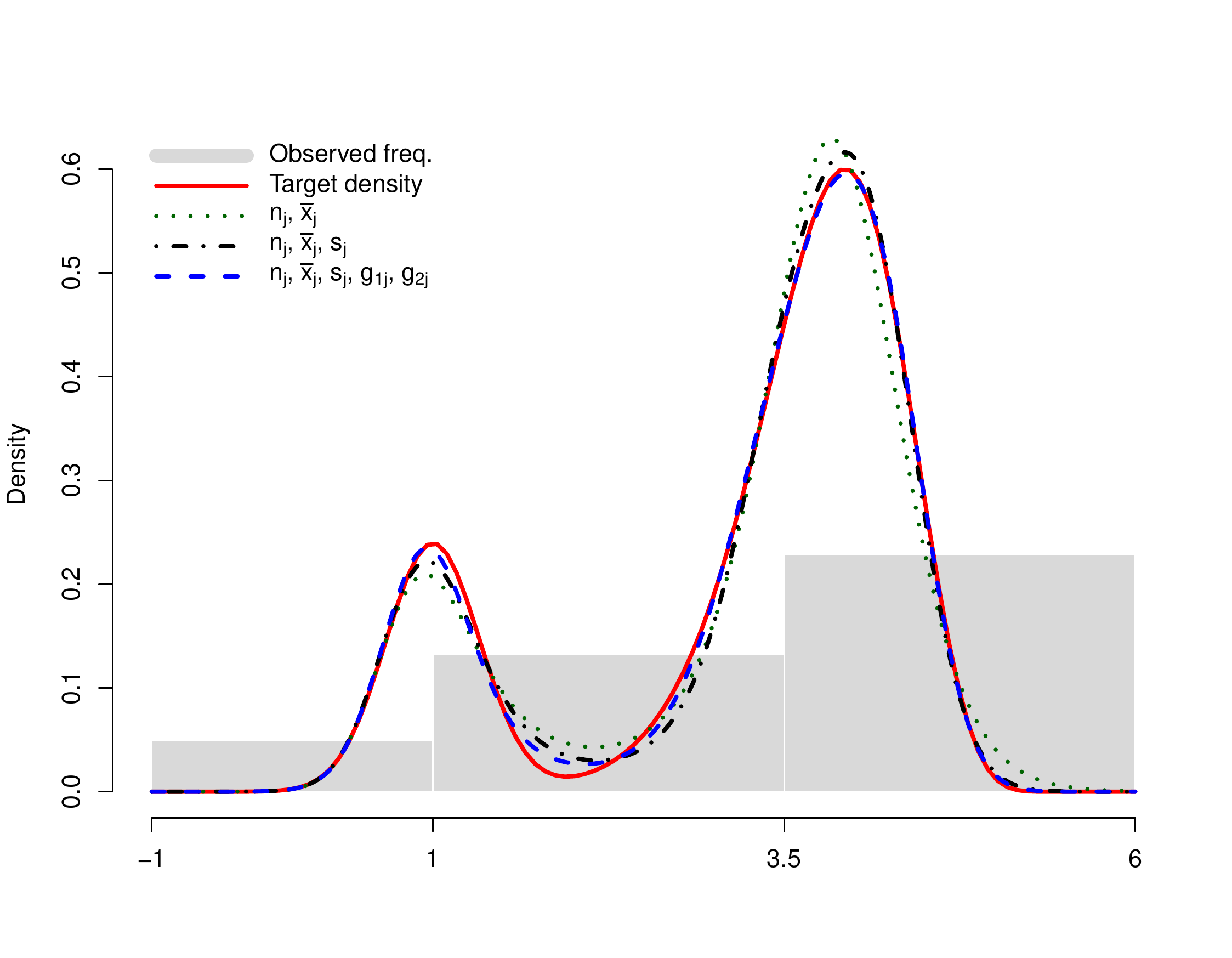}
\caption{Simulation study ($n=1000$, $J=3$) -- Density estimates from
  tabulated summary statistics $\D_1$ (dotted), $\D_2$ (dot-dashed) and
  $\D_4$ (dashed) averaged over the $S=500$ replicates, with the
  'true' underlying density (solid curve).}
\label{Simulation:n1000:J3:Fig}
\end{figure}

\begin{figure}
  \centering
  \begin{tabular}{cc}
    \includegraphics[width=.49\textwidth]{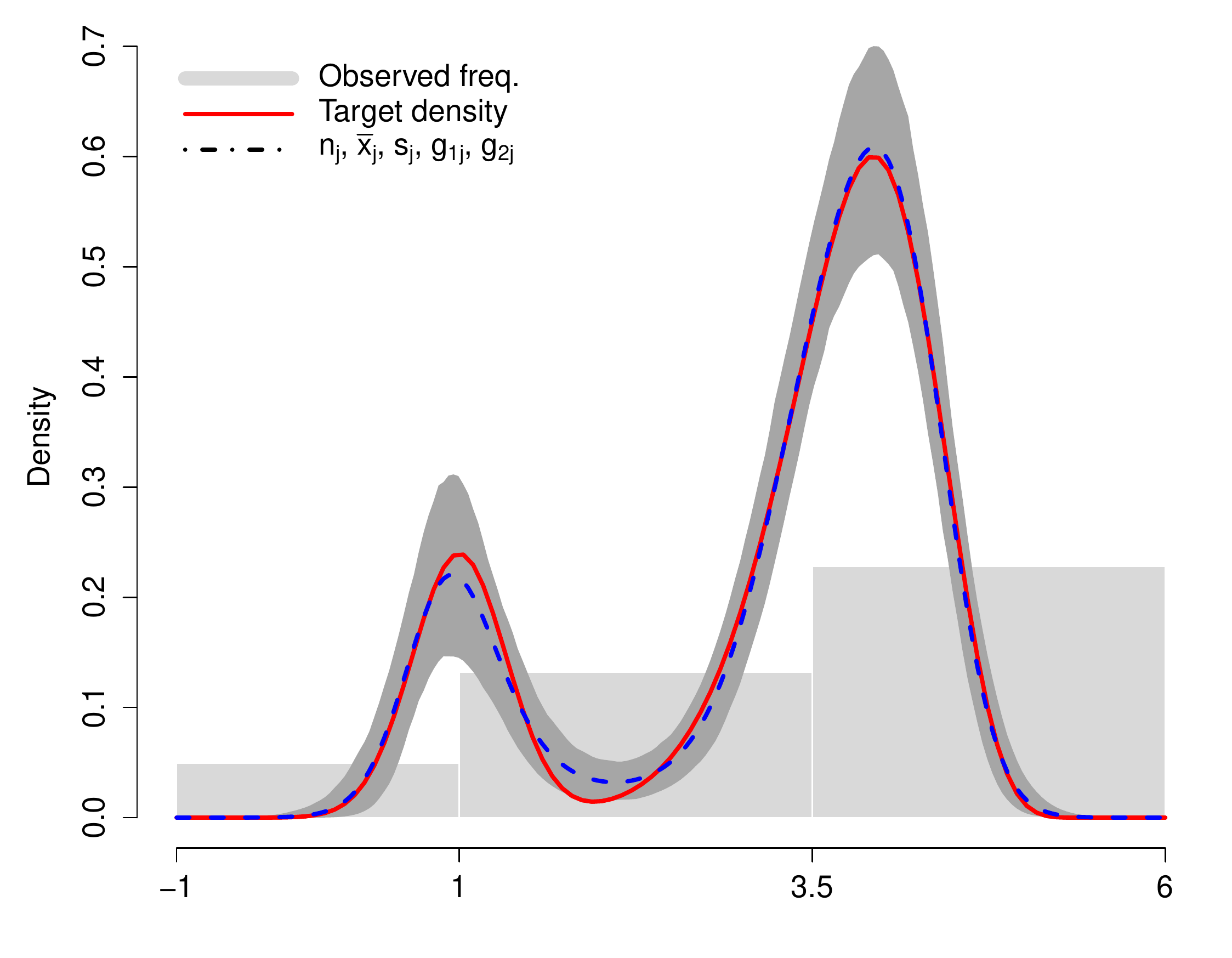} & \includegraphics[width=.49\textwidth]{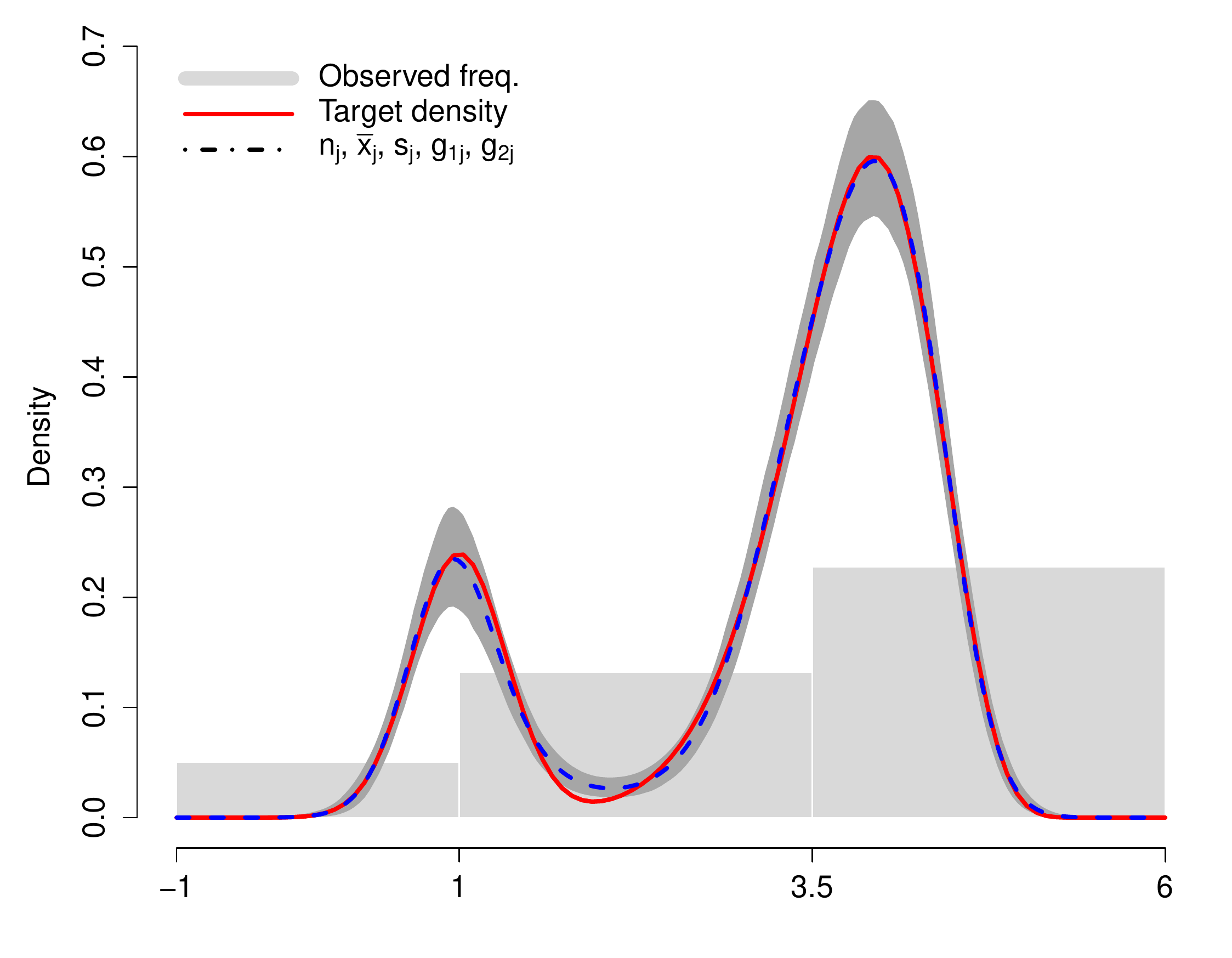}
  \end{tabular}

  \caption{Simulation study -- Connected pointwise intervals containing 95\% of the estimated densities over the $S=500$ datasets
    and obtained using tabulated summary statistics ${\cal D}_4$ over $J=3$ classes. Left panel: $n=250$ ; Right panel: $n=1\,000$.}
\label{Simulation:EstimatedDensities:J3:Fig}
\end{figure}

In this section, we evaluate the performances of the estimation method
proposed in Section \ref{Methodology:Sec} by means of extensive
simulations.  Independent and identically distributed data were
generated from a mixture density,
$$f(x) = w_1 f_1(x) + w_2 f_2(5.6-x)$$
where $f_1(\cdot)$ corresponds to a Normal density with mean $1.0$ and
variance $9.0^{-1}$, $f_2(\cdot)$ to a Gamma density with mean $11/6$
and variance $11/6^2$, weighted respectively by $w_1=.20$ and
$w_2=.80$.  It corresponds to the solid red curve in
Figs.\,\ref{Simulation:n1000:J3:Fig} and
\ref{Simulation:EstimatedDensities:J3:Fig} that could be viewed as the
underlying distribution of log transformed positive data.  Datasets of
size $n=250$, $1\,000$ or $3\,000$ were generated $S=500$ times and
grouped into either $J=3$ or $J=5$ partitioning classes
$\{{\cal C}_j:j=1,\ldots,J\}$ with interval extremities given by
$\{-1.0,1.0,3.5,6.0\}$ and $\{-1.0,1.0,2.2,3.5,4.8,6.0\}$,
respectively. Tabulated frequencies $(n_j)_{J=1}^J$ and the associated
local central moments $(m_{rj})_{j=1}^J$ of order $r=1$ up to $4$ were
computed and used with the methodology of Section
\ref{Methodology:Sec} to produce an estimate
$\hat{f}(\cdot|{\cal D}_r)$ of the underlying density on $(-1.0,6.0)$,
with ${\cal D}_r=\bigcup_{j=1}^J\{n_j,m_{1j},\ldots,m_{rj}\}$ denoting
the available data. Selected quantile estimates
$\hat{Q}(p|{\cal D}_r)$ were computed using that density and compared
to the `true' quantile values associated to $f(\cdot)$. Biases,
standard deviations (SD), root mean squared errors (RMSE) and
effective coverages of 95\% and 90\% credible intervals are given in
Tables \ref{Quantiles:n250:Tab}, \ref{Quantiles:n1000:Tab} and
\ref{Quantiles:n3000:Tab} for different samples sizes and number of
classes. As expected, biases for the point estimator of a given
quantile tend to decrease with the sample size and the number of classes
for which tabulated summary statistics are observed. They are
already very small when $n=250$ with $r=4$ moments reported in only
$J=3$ classes, see Fig.\,\ref{Simulation:EstimatedDensities:J3:Fig}
for a graphical illustration at the density level when $n=250$ and
$n=1000$. An exception concerns the 20\% quantile ($=1.793$) that is
not so well estimated whatever the simulation setting: it corresponds
to the region surrounding the local minimum of the mixture
density between the two modes. Increasing the number of reported
central moments tends to improve the estimation of density and
quantiles, with 4 moments being preferable, see
Fig.\,\ref{Simulation:n1000:J3:Fig} for an evolution of the averaged
density estimates (over the $S=500$ replicates) starting with ${\D_1}$
(dotted curve) to $\D_4$ (dashed curve) when $n=1\,000$. This is a
remarkable improvement over the estimate that would be obtained using
only observed frequencies. Whatever the sample size, the effective
coverages of 90\% and 95\% credible intervals for the reported
quantiles (except the 20\% one) are in agreement with their nominal
values when 4 central moments ($\D_4$) are reported. Moderate
undercoverages can be observed with $\D_2$, while the effective
coverages can be larger than expected when just the means are reported
in addition to frequencies ($\D_1$).  Global metrics were also
calculated to compare the true and estimated quantile functions,
\begin{align*}
  \ell_1(Q,\hat{Q}) = \int_0^1 \big|Q(p|\hat\thetavec)-Q(p)\big|\,dp\,
\end{align*}
as well the true and estimated densities,
\begin{align*}
   \text{RIMSE}(f,\hat{f}) &= \int_\rit \big(f(x|\hat\thetavec)-f(x)\big)^2\,f(x)\,dx~~;\\
   \text{KL}(f,\hat{f}) &= \int_\rit f(x) \log\left({f(x)\over\hat{f}(x)}\right)\,dx~,
\end{align*}
see Table \ref{L1Q-RIMSE:Tab} for their median values over the $S=500$
simulated datasets with tabulated summary statistics $\D_r$
($r=1,2,4$) in 3 or 5 classes. This suggests that the extra
information provided by additional central moments for quantile or
density estimation is even more valuable when the number of classes
is small.

\begin{table}[ht]
  \centering
  % \resizebox{\textwidth}{!}
  {
    \begin{tabular}{rrrrrrrrrrrr}
      \toprule %\hline
      $p$     & 0.1   & 0.2   & 0.3   & 0.4   & 0.5   & 0.6   & 0.7   & 0.8   & 0.9   \\ 
      $Q(p)$  & 1.000 & 1.793 & 3.122 & 3.430 & 3.643 & 3.822 & 3.989 & 4.163 & 4.375 \\
      \midrule
      \multicolumn{10}{c}{$J=3$ (classes)}\\
      \midrule
      $\hat{Q}(p|{\cal D}_1)$
              & 1.084 & 2.071 & 3.081 & 3.421 & 3.630 & 3.801 & 3.965 & 4.144 & 4.383 \\           
      Bias    & 0.084 & 0.278 &-0.041 &-0.009 &-0.014 &-0.021 &-0.024 &-0.019 & 0.007 \\  
      SD      & 0.121 & 0.401 & 0.154 & 0.074 & 0.054 & 0.045 & 0.040 & 0.038 & 0.048 \\         
      RMSE    & 0.147 & 0.488 & 0.160 & 0.075 & 0.056 & 0.050 & 0.046 & 0.043 & 0.049 \\         
      95\% CI & 0.950 & 0.694 & 0.930 & 0.964 & 0.958 & 0.964 & 0.968 & 0.934 & 0.926 \\
      90\% CI & 0.904 & 0.622 & 0.880 & 0.922 & 0.924 & 0.926 & 0.920 & 0.870 & 0.866 \\
      \midrule
      $\hat{Q}(p|{\cal D}_2)$
              & 1.016 & 2.003 & 3.166 & 3.466 & 3.663 & 3.830 & 3.988 & 4.155 & 4.363 \\       
      Bias    & 0.016 & 0.210 & 0.045 & 0.036 & 0.020 & 0.008 &-0.001 &-0.009 &-0.013 \\ 
      SD      & 0.082 & 0.493 & 0.124 & 0.064 & 0.050 & 0.044 & 0.040 & 0.039 & 0.040 \\       
      RMSE    & 0.084 & 0.536 & 0.132 & 0.073 & 0.054 & 0.044 & 0.040 & 0.040 & 0.042 \\       
      95\% CI & 0.950 & 0.656 & 0.860 & 0.886 & 0.908 & 0.932 & 0.934 & 0.896 & 0.840 \\
      90\% CI & 0.906 & 0.586 & 0.778 & 0.830 & 0.842 & 0.878 & 0.876 & 0.848 & 0.762 \\
      \midrule      
      $\hat{Q}(p|{\cal D}_4)$
              & 1.007 & 2.004 & 3.121 & 3.435 & 3.646 & 3.823 & 3.988 & 4.160 & 4.369 \\    
      Bias    & 0.007 & 0.211 &-0.001 & 0.005 & 0.003 & 0.001 &-0.001 &-0.004 &-0.006 \\  
      SD      & 0.081 & 0.480 & 0.121 & 0.068 & 0.053 & 0.046 & 0.041 & 0.040 & 0.041 \\    
      RMSE    & 0.082 & 0.525 & 0.121 & 0.068 & 0.054 & 0.046 & 0.041 & 0.040 & 0.041 \\    
      95\% CI & 0.948 & 0.740 & 0.946 & 0.954 & 0.950 & 0.948 & 0.962 & 0.950 & 0.948 \\    
      90\% CI & 0.910 & 0.694 & 0.918 & 0.912 & 0.908 & 0.912 & 0.924 & 0.904 & 0.880 \\    
      \midrule
      \multicolumn{10}{c}{$J=5$ (classes)}\\
      \midrule
      $\hat{Q}(p|{\cal D}_1)$
              & 0.986 & 1.920 & 3.104 & 3.429 & 3.648 & 3.827 & 3.992 & 4.159 & 4.364 \\  
      Bias    &-0.014 & 0.127 &-0.018 &-0.001 & 0.005 & 0.006 & 0.002 &-0.004 &-0.011 \\ 
      SD      & 0.081 & 0.532 & 0.129 & 0.074 & 0.057 & 0.048 & 0.042 & 0.041 & 0.044 \\ 
      RMSE    & 0.082 & 0.547 & 0.131 & 0.074 & 0.058 & 0.048 & 0.043 & 0.042 & 0.046 \\ 
      95\% CI & 0.910 & 0.650 & 0.938 & 0.930 & 0.936 & 0.932 & 0.912 & 0.882 & 0.898 \\ 
      90\% CI & 0.866 & 0.598 & 0.900 & 0.894 & 0.892 & 0.892 & 0.864 & 0.818 & 0.838 \\ 
      \midrule
      $\hat{Q}(p|{\cal D}_2)$
              & 0.998 & 1.914 & 3.100 & 3.423 & 3.640 & 3.820 & 3.988 & 4.162 & 4.373 \\ 
      Bias    &-0.002 & 0.121 &-0.022 &-0.007 &-0.003 &-0.002 &-0.001 &-0.002 &-0.002 \\ 
      SD      & 0.081 & 0.532 & 0.132 & 0.073 & 0.056 & 0.048 & 0.043 & 0.043 & 0.047 \\ 
      RMSE    & 0.081 & 0.545 & 0.134 & 0.073 & 0.056 & 0.048 & 0.043 & 0.043 & 0.047 \\ 
      95\% CI & 0.920 & 0.608 & 0.948 & 0.934 & 0.934 & 0.924 & 0.910 & 0.842 & 0.732 \\ 
      90\% CI & 0.878 & 0.548 & 0.904 & 0.880 & 0.884 & 0.878 & 0.860 & 0.792 & 0.654 \\ 
      \midrule      
      $\hat{Q}(p|{\cal D}_4)$
              & 1.012 & 2.001 & 3.120 & 3.435 & 3.646 & 3.822 & 3.987 & 4.157 & 4.368 \\
      Bias    & 0.012 & 0.208 &-0.002 & 0.005 & 0.002 & 0.000 &-0.003 &-0.006 &-0.008 \\
      SD      & 0.091 & 0.480 & 0.120 & 0.068 & 0.054 & 0.046 & 0.041 & 0.039 & 0.040 \\
      RMSE    & 0.092 & 0.523 & 0.120 & 0.068 & 0.054 & 0.046 & 0.041 & 0.040 & 0.041 \\
      95\% CI & 0.952 & 0.728 & 0.946 & 0.952 & 0.950 & 0.948 & 0.960 & 0.952 & 0.940 \\
      90\% CI & 0.914 & 0.680 & 0.928 & 0.906 & 0.910 & 0.906 & 0.922 & 0.910 & 0.874 \\
      \bottomrule
    \end{tabular}}
  \caption{Simulation study ($n=250$) -- Selected $p-$quantile estimates using
    tabulated summary statistics in ${\cal D}_r$ with $r=1,2,4$: bias,
    standard deviation, root mean squared error (RMSE) and effective
    coverages of 95\% and 90\% credible intervals (based on $S=500$ replicates).}
  \label{Quantiles:n250:Tab}
\end{table}

\begin{table}[ht]
  \centering
  \resizebox{\textwidth}{!}
  {
    \begin{tabular}{rrrrrrrrrrrrr}
      \toprule %\hline
      $p$     & 0.1   & 0.2   & 0.3   & 0.4   & 0.5   & 0.6   & 0.7   & 0.8   & 0.9   & 0.95 \\ 
      $Q(p)$  & 1.000 & 1.793 & 3.122 & 3.430 & 3.643 & 3.822 & 3.989 & 4.163 & 4.375 & 4.530 \\ 
      \midrule
      \multicolumn{11}{c}{$J=3$ (classes)}\\
      \midrule
      $\hat{Q}(p|{\cal D}_1)$
              & 1.026 & 1.940 & 3.121 & 3.436 & 3.634 & 3.800 & 3.960 & 4.138 & 4.381 & 4.586 \\ 
      Bias    & 0.027 & 0.147 &-0.001 & 0.005 &-0.009 &-0.022 &-0.029 &-0.025 & 0.006 & 0.056 \\ 
      SD      & 0.047 & 0.257 & 0.069 & 0.037 & 0.028 & 0.024 & 0.021 & 0.020 & 0.025 & 0.033 \\ 
      RMSE    & 0.054 & 0.297 & 0.069 & 0.037 & 0.030 & 0.032 & 0.036 & 0.032 & 0.025 & 0.065 \\ 
      95\% CI & 0.944 & 0.836 & 0.954 & 0.930 & 0.954 & 0.972 & 0.980 & 0.948 & 0.960 & 0.998 \\ 
      90\% CI & 0.878 & 0.786 & 0.910 & 0.894 & 0.916 & 0.920 & 0.920 & 0.892 & 0.918 & 0.978 \\ 
      \midrule
      $\hat{Q}(p|{\cal D}_2)$
              & 1.005 & 1.867 & 3.159 & 3.452 & 3.653 & 3.826 & 3.989 & 4.159 & 4.369 & 4.527 \\ 
      Bias    & 0.006 & 0.074 & 0.037 & 0.022 & 0.010 & 0.004 & 0.000 &-0.005 &-0.007 &-0.003 \\ 
      SD      & 0.040 & 0.302 & 0.058 & 0.035 & 0.028 & 0.025 & 0.023 & 0.021 & 0.021 & 0.023 \\ 
      RMSE    & 0.040 & 0.311 & 0.068 & 0.042 & 0.030 & 0.025 & 0.023 & 0.022 & 0.022 & 0.023 \\ 
      95\% CI & 0.960 & 0.752 & 0.834 & 0.880 & 0.918 & 0.916 & 0.910 & 0.886 & 0.874 & 0.816 \\ 
      90\% CI & 0.902 & 0.698 & 0.768 & 0.812 & 0.844 & 0.866 & 0.844 & 0.820 & 0.788 & 0.728 \\ 
      \midrule      
      $\hat{Q}(p|{\cal D}_4)$
              & 0.993 & 1.904 & 3.125 & 3.429 & 3.642 & 3.822 & 3.991 & 4.165 & 4.376 & 4.530 \\ 
      Bias    &-0.006 & 0.111 & 0.003 &-0.001 &-0.001 & 0.000 & 0.002 & 0.002 & 0.001 & 0.001 \\ 
      SD      & 0.037 & 0.331 & 0.058 & 0.037 & 0.029 & 0.025 & 0.023 & 0.021 & 0.021 & 0.023 \\ 
      RMSE    & 0.037 & 0.349 & 0.058 & 0.037 & 0.029 & 0.025 & 0.023 & 0.022 & 0.021 & 0.023 \\ 
      95\% CI & 0.950 & 0.816 & 0.938 & 0.942 & 0.940 & 0.946 & 0.948 & 0.956 & 0.952 & 0.958 \\ 
      90\% CI & 0.904 & 0.776 & 0.892 & 0.898 & 0.890 & 0.904 & 0.898 & 0.896 & 0.904 & 0.908 \\ 
      \midrule
      \multicolumn{11}{c}{$J=5$ (classes)}\\
      \midrule
      $\hat{Q}(p|{\cal D}_1)$
              & 0.989 & 1.864 & 3.108 & 3.427 & 3.648 & 3.828 & 3.992 & 4.161 & 4.367 & 4.522 \\
      Bias    &-0.010 & 0.071 &-0.014 &-0.003 & 0.005 & 0.006 & 0.003 &-0.003 &-0.009 &-0.008 \\
      SD      & 0.034 & 0.370 & 0.061 & 0.039 & 0.032 & 0.026 & 0.023 & 0.022 & 0.023 & 0.025 \\
      RMSE    & 0.036 & 0.377 & 0.063 & 0.040 & 0.032 & 0.027 & 0.023 & 0.022 & 0.024 & 0.027 \\
      95\% CI & 0.950 & 0.734 & 0.920 & 0.928 & 0.936 & 0.930 & 0.916 & 0.858 & 0.934 & 0.966 \\
      90\% CI & 0.900 & 0.686 & 0.848 & 0.894 & 0.884 & 0.880 & 0.830 & 0.786 & 0.882 & 0.928 \\
      \midrule
      $\hat{Q}(p|{\cal D}_2)$
              & 0.992 & 1.836 & 3.106 & 3.426 & 3.642 & 3.821 & 3.989 & 4.164 & 4.378 & 4.536 \\ 
      Bias    &-0.008 & 0.043 &-0.016 &-0.004 &-0.001 &-0.001 &-0.001 & 0.001 & 0.003 & 0.006 \\ 
      SD      & 0.036 & 0.357 & 0.064 & 0.038 & 0.030 & 0.026 & 0.024 & 0.023 & 0.024 & 0.026 \\ 
      RMSE    & 0.037 & 0.360 & 0.066 & 0.038 & 0.030 & 0.026 & 0.024 & 0.023 & 0.024 & 0.026 \\ 
      95\% CI & 0.942 & 0.730 & 0.912 & 0.938 & 0.932 & 0.924 & 0.890 & 0.816 & 0.692 & 0.808 \\ 
      90\% CI & 0.874 & 0.690 & 0.842 & 0.890 & 0.892 & 0.866 & 0.820 & 0.744 & 0.592 & 0.734 \\ 
      \midrule      
      $\hat{Q}(p|{\cal D}_4)$
              & 0.995 & 1.914 & 3.121 & 3.429 & 3.643 & 3.823 & 3.990 & 4.164 & 4.374 & 4.530 \\ 
      Bias    &-0.005 & 0.121 &-0.001 &-0.001 & 0.000 & 0.001 & 0.001 & 0.000 &-0.002 & 0.000 \\ 
      SD      & 0.037 & 0.342 & 0.059 & 0.037 & 0.029 & 0.025 & 0.023 & 0.022 & 0.022 & 0.023 \\ 
      RMSE    & 0.037 & 0.363 & 0.059 & 0.037 & 0.029 & 0.025 & 0.023 & 0.022 & 0.022 & 0.023 \\ 
      95\% CI & 0.954 & 0.804 & 0.936 & 0.942 & 0.940 & 0.940 & 0.946 & 0.956 & 0.944 & 0.956 \\ 
      90\% CI & 0.910 & 0.754 & 0.890 & 0.900 & 0.896 & 0.904 & 0.900 & 0.880 & 0.892 & 0.914 \\ 
      \bottomrule
    \end{tabular}}
  \caption{Simulation study ($n=1000$) -- Selected $p-$quantile estimates using
    tabulated summary statistics in ${\cal D}_r$ with $r=1,2,4$: bias,
    standard deviation, root mean squared error (RMSE) and effective
    coverages of 95\% and 90\% credible intervals (based on $S=500$ replicates).}
  \label{Quantiles:n1000:Tab}
\end{table}

\begin{table}[ht]
  \centering
  \resizebox{\textwidth}{!}
  {
    \begin{tabular}{rrrrrrrrrrrrr}
      \toprule %\hline
      $p$     & 0.1   & 0.2   & 0.3   & 0.4   & 0.5   & 0.6   & 0.7   & 0.8   & 0.9   & 0.95  & 0.99  \\ 
      $Q(p)$  & 1.000 & 1.793 & 3.122 & 3.430 & 3.643 & 3.822 & 3.989 & 4.163 & 4.375 & 4.530 & 4.778 \\ 
      \midrule
      \multicolumn{12}{c}{$J=3$ (classes)}\\
      \midrule
      $\hat{Q}(p|{\cal D}_1)$
              & 1.013 & 1.897 & 3.127 & 3.438 & 3.634 & 3.796 & 3.954 & 4.131 & 4.376 & 4.591 & 5.052 \\ 
      Bias    & 0.013 & 0.104 & 0.005 & 0.008 &-0.010 &-0.025 &-0.035 &-0.033 & 0.000 & 0.061 & 0.274 \\ 
      SD      & 0.025 & 0.145 & 0.038 & 0.021 & 0.016 & 0.013 & 0.011 & 0.010 & 0.013 & 0.020 & 0.045 \\ 
      RMSE    & 0.029 & 0.179 & 0.038 & 0.023 & 0.019 & 0.029 & 0.037 & 0.034 & 0.013 & 0.064 & 0.278 \\ 
      95\% CI & 0.948 & 0.934 & 0.990 & 0.928 & 0.960 & 0.992 & 0.996 & 0.994 & 0.996 & 1.000 & 1.000 \\ 
      90\% CI & 0.886 & 0.874 & 0.972 & 0.866 & 0.916 & 0.944 & 0.974 & 0.968 & 0.996 & 0.996 & 1.000 \\ 
      \midrule
      $\hat{Q}(p|{\cal D}_2)$
              & 1.004 & 1.810 & 3.153 & 3.444 & 3.649 & 3.825 & 3.990 & 4.160 & 4.368 & 4.524 & 4.797 \\ 
      Bias    & 0.005 & 0.017 & 0.032 & 0.014 & 0.005 & 0.003 & 0.001 &-0.003 &-0.007 &-0.006 & 0.019 \\ 
      SD      & 0.023 & 0.170 & 0.032 & 0.021 & 0.017 & 0.014 & 0.013 & 0.012 & 0.011 & 0.012 & 0.021 \\ 
      RMSE    & 0.023 & 0.171 & 0.045 & 0.025 & 0.018 & 0.014 & 0.013 & 0.013 & 0.014 & 0.013 & 0.028 \\ 
      95\% CI & 0.954 & 0.828 & 0.762 & 0.868 & 0.936 & 0.942 & 0.904 & 0.932 & 0.886 & 0.818 & 0.942 \\ 
      90\% CI & 0.910 & 0.762 & 0.686 & 0.794 & 0.866 & 0.876 & 0.840 & 0.832 & 0.810 & 0.750 & 0.866 \\ 
      \midrule      
      $\hat{Q}(p|{\cal D}_4)$
              & 0.994 & 1.868 & 3.126 & 3.429 & 3.643 & 3.822 & 3.991 & 4.165 & 4.376 & 4.528 & 4.782 \\ 
      Bias    &-0.006 & 0.075 & 0.004 &-0.001 &-0.000 & 0.001 & 0.001 & 0.002 & 0.000 &-0.002 & 0.004 \\ 
      SD      & 0.021 & 0.223 & 0.033 & 0.021 & 0.017 & 0.014 & 0.013 & 0.012 & 0.012 & 0.012 & 0.018 \\ 
      RMSE    & 0.022 & 0.235 & 0.033 & 0.022 & 0.017 & 0.014 & 0.013 & 0.012 & 0.012 & 0.013 & 0.018 \\ 
      95\% CI & 0.948 & 0.904 & 0.946 & 0.948 & 0.956 & 0.956 & 0.954 & 0.946 & 0.966 & 0.964 & 0.968 \\ 
      90\% CI & 0.902 & 0.850 & 0.894 & 0.880 & 0.894 & 0.914 & 0.906 & 0.910 & 0.914 & 0.918 & 0.918 \\ 
      \midrule
      \multicolumn{12}{c}{$J=5$ (classes)}\\
      \midrule
      $\hat{Q}(p|{\cal D}_1)$
              & 0.999 & 1.864 & 3.115 & 3.430 & 3.644 & 3.820 & 3.986 & 4.161 & 4.376 & 4.530 & 4.771 \\ 
      Bias    &-0.001 & 0.071 &-0.007 & 0.000 & 0.001 &-0.002 &-0.004 &-0.002 & 0.001 & 0.000 &-0.007 \\ 
      SD      & 0.020 & 0.258 & 0.034 & 0.023 & 0.018 & 0.015 & 0.014 & 0.013 & 0.017 & 0.019 & 0.019 \\ 
      RMSE    & 0.020 & 0.268 & 0.035 & 0.023 & 0.018 & 0.015 & 0.014 & 0.013 & 0.017 & 0.019 & 0.020 \\ 
      95\% CI & 0.958 & 0.846 & 0.944 & 0.952 & 0.976 & 0.982 & 0.938 & 0.908 & 0.974 & 0.976 & 0.984 \\ 
      90\% CI & 0.920 & 0.772 & 0.882 & 0.880 & 0.922 & 0.946 & 0.858 & 0.826 & 0.926 & 0.932 & 0.934 \\ 
      \midrule
      $\hat{Q}(p|{\cal D}_2)$
              & 0.997 & 1.831 & 3.114 & 3.433 & 3.646 & 3.822 & 3.987 & 4.161 & 4.375 & 4.530 & 4.777 \\ 
      Bias    &-0.003 & 0.038 &-0.008 & 0.003 & 0.003 & 0.000 &-0.002 &-0.002 &-0.001 & 0.000 &-0.001 \\ 
      SD      & 0.021 & 0.252 & 0.035 & 0.022 & 0.017 & 0.014 & 0.013 & 0.013 & 0.013 & 0.013 & 0.018 \\ 
      RMSE    & 0.022 & 0.255 & 0.036 & 0.022 & 0.017 & 0.014 & 0.014 & 0.013 & 0.013 & 0.013 & 0.018 \\ 
      95\% CI & 0.942 & 0.814 & 0.934 & 0.932 & 0.938 & 0.910 & 0.896 & 0.852 & 0.746 & 0.894 & 0.972 \\ 
      90\% CI & 0.902 & 0.750 & 0.860 & 0.882 & 0.876 & 0.858 & 0.800 & 0.788 & 0.668 & 0.824 & 0.924 \\ 
      \midrule      
      $\hat{Q}(p|{\cal D}_4)$
              & 0.996 & 1.875 & 3.123 & 3.430 & 3.644 & 3.823 & 3.990 & 4.164 & 4.375 & 4.527 & 4.778 \\ 
      Bias    &-0.004 & 0.082 & 0.001 & 0.000 & 0.000 & 0.001 & 0.001 & 0.001 &-0.001 &-0.003 & 0.000 \\ 
      SD      & 0.021 & 0.241 & 0.033 & 0.021 & 0.017 & 0.014 & 0.013 & 0.012 & 0.012 & 0.013 & 0.016 \\ 
      RMSE    & 0.021 & 0.255 & 0.033 & 0.021 & 0.017 & 0.014 & 0.013 & 0.012 & 0.012 & 0.013 & 0.016 \\ 
      95\% CI & 0.958 & 0.860 & 0.958 & 0.946 & 0.954 & 0.960 & 0.948 & 0.942 & 0.962 & 0.958 & 0.982 \\ 
      90\% CI & 0.902 & 0.820 & 0.896 & 0.884 & 0.892 & 0.910 & 0.904 & 0.904 & 0.904 & 0.928 & 0.952 \\ 
      \bottomrule
    \end{tabular}}
  \caption{Simulation study ($n=3000$) -- Selected $p-$quantile estimates using
    tabulated summary statistics in ${\cal D}_r$ with $r=1,2,4$: bias,
    standard deviation, root mean squared error (RMSE) and effective
    coverages of 95\% and 90\% credible intervals (based on $S=500$ replicates).}
  \label{Quantiles:n3000:Tab}
\end{table}

\begin{table}[ht]
  \centering
  % \resizebox{\textwidth}{!}
  {
    \begin{tabular}{rlccccccc}
\toprule
      & &\multicolumn{3}{c}{$J=3$ (classes)}      && \multicolumn{3}{c}{$J=5$ (classes)} \\
      \cline{3-5}  \cline{7-9} 
      $n$      &Metric &${\cal D}_1$ & ${\cal D}_2$ & ${\cal D}_4$ && ${\cal D}_1$ & ${\cal D}_2$ & ${\cal D}_4$ \\
      \midrule
      $250$    & $\ell_1(Q,\hat{Q})$ & 0.087 & 0.072 & 0.067  && 0.072 & 0.071 & 0.068 \\ % L1Q  
               & RIMSE & 0.048 & 0.044 & 0.037  && 0.043 & 0.043 & 0.037 \\ % RIMSE
               & K-L   & 0.034 & 0.022 & 0.016  && 0.018 & 0.019 & 0.016 \\ % K-L        
      $1\,000$ & $\ell_1(Q,\hat{Q})$ & 0.051 & 0.038 & 0.034  && 0.037 & 0.036 & 0.034 \\ % L1Q
               & RIMSE & 0.041 & 0.025 & 0.020  && 0.025 & 0.024 & 0.020 \\ % RIMSE
               & K-L   & 0.025 & 0.011 & 0.006  && 0.006 & 0.006 & 0.005 \\ % K-L      
      $3\,000$ & $\ell_1(Q,\hat{Q})$ & 0.044 & 0.025 & 0.021  && 0.023 & 0.021 & 0.021 \\ % L1Q  
               & RIMSE & 0.044 & 0.018 & 0.014  && 0.018 & 0.016 & 0.013 \\ % RIMSE
               & K-L   & 0.026 & 0.008 & 0.003  && 0.003 & 0.002 & 0.002 \\ % K-L  
      \bottomrule
    \end{tabular}
    \caption{Simulation study -- $L_1-$distance, $\ell_1(Q,\hat{Q})$,
      between the true and the estimated quantile functions ; 
      Root integrated mean squared error (RIMSE)
      and Kullback-Liebler divergence (K-L)
      comparing the true and
      estimated density functions. Median values of these metrics
      (over $S=500$ simulated datasets of size $n$) are reported with
      estimation performed from tabulated summary statistics
      ${\cal D}_r$ (with $r=1,2,4$) in $J$ classes.}
  }
  \label{L1Q-RIMSE:Tab}
\end{table}

In summary, this simulation study confirms the added value of central
moments over isolated frequencies for density estimation. With only 15
numbers (the frequency and the 4 central moments in each of the $J=3$
classes), an accurate and precise density estimate could be obtained
from summary statistics in just $3$ classes, see
Fig.\,\ref{Simulation:EstimatedDensities:J3:Fig} for a graphical
representation. The simulation study also suggests that this method
can be used to estimate quantiles and, consequently, values at risk
(VaR) in a reliable way.

\section{Application}
\begin{figure}
   \centering
{\includegraphics[width=14.5cm]{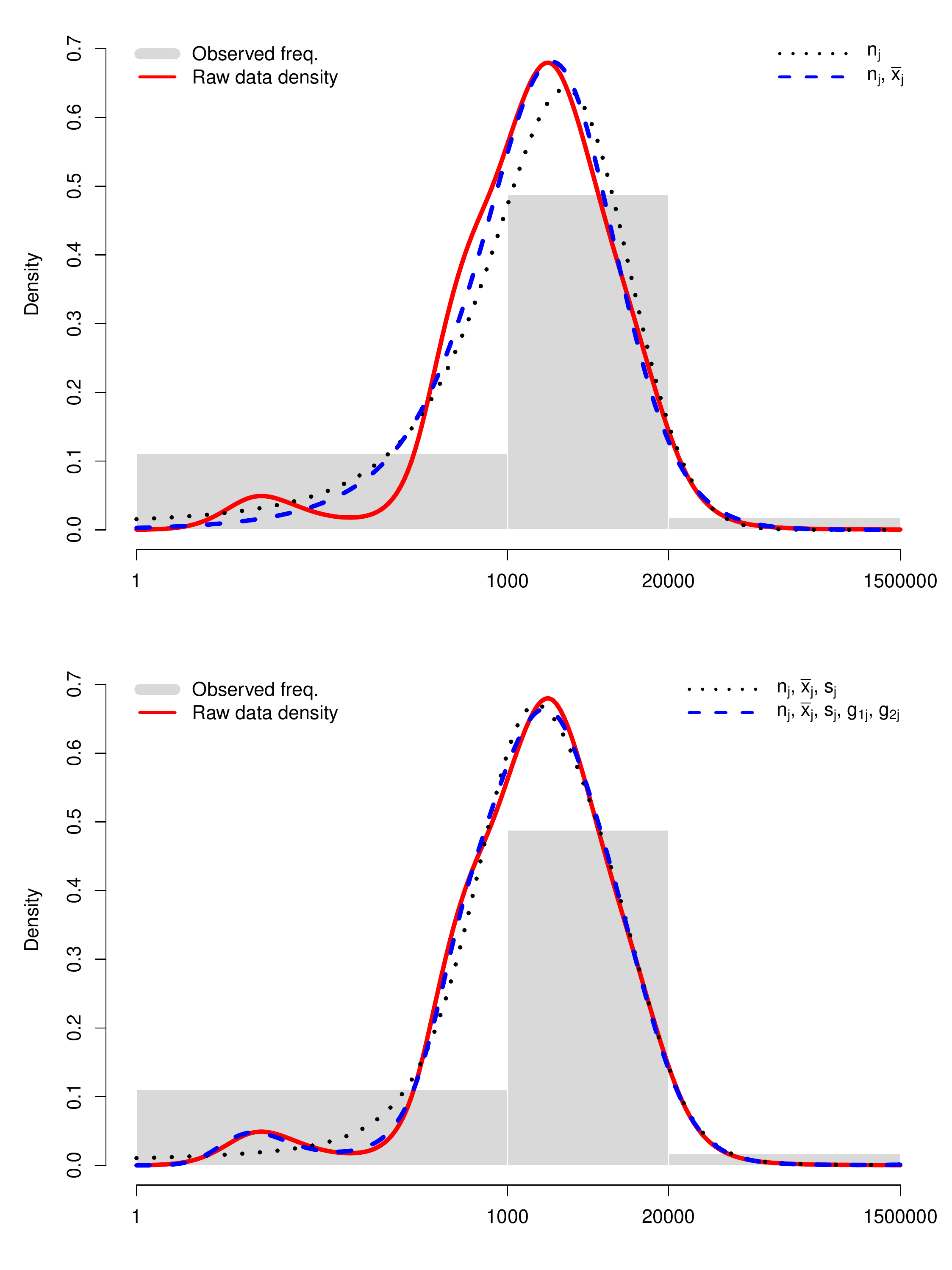}}
\caption{Car insurance data: observed frequencies and density
  estimates (on the $\log_{10}$-scale) using tabulated sample moments.}
\label{PaymentData:Fit:Fig}
\end{figure}

Table \ref{Payment:Tab} provides summary statistics on insurance claim
amount data (in euros). For confidentiality reasons, the 3\,518 data
were rescaled and gathered in $J=3$ classes of increasing width. Besides the class
frequencies $n_j$, the sample mean, standard deviation, skewness and
kurtosis of the $\log_{10}$ transformed claims within each class are also
provided.
\begin{table}
\begin{center}
  \begin{tabular}{lrcccSS}
  \toprule
                          &  &     \multicolumn{5}{c}{$\log_{10}$(Claim)}\\
\cline{3-7}
Claim                &\mcc{Freq.}& Interval         & \mcc{Mean}             & Std.dev & \mcc{Skewness} & \mcc{Kurtosis}\\
Interval                 &{$n_j$}& $\mathcal{C}_j$  & \mcc{$\overline{x}_j$} & $s_j$    & \mcc{$g_{1j}$} & \mcc{$g_{2j}$} \\
\midrule
(1\,;\,1\,000]            & 1168 & (0.00\,;\,3.00]  & 2.462 & 0.580 & -1.793 &  2.401 \\
(1\,000\,;\,20\,000]      & 2234 & (3.00\,;\,4.30]  & 3.529 & 0.336 &  0.375 & -0.836 \\
(20\,000\,;\,1\,500\,000] &  116 & (4.30\,;\,6.18]  & 4.556 & 0.275 &  2.603 &  9.416 \\
\bottomrule
\end{tabular}
\caption{Car insurance data: summary statistics for $n=3518$ grouped claim data (in euros).}\label{Payment:Tab}
\end{center}
\end{table}
Figure \ref{PaymentData:Fit:Fig} displays the histogram corresponding
to the grouped data frequencies. The thick solid (red) line
corresponds to the 'target' density estimated from the precise
individual ($\log_{10}$) claims shared with us in confidentiality by
the insurance company.  Estimation of the density using the grouped
summary statistics in Table \ref{Payment:Tab} was performed using the
methods described in Sections
\ref{DensityEstimationFromFreqUsingEM:Sec} and
\ref{DensityEstimationFromMomentsUsingEM:Sec} with $K=25$ B-splines
associated to equidistant knots on $(0.0,6.18)$.  Computation was
performed in less than one second using the R-package \texttt{degross}
(Density Estimation from GRouped Summary Statistics) developed and
maintained by the author.  The top graph in the figure compares the
'target' density with the density estimates obtained from the grouped
data frequencies ($\D_0$, dotted line) and from the addition of the
grouped sample means ($\D_1$, dashed line). The bottom graph further
considers the cumulative addition of the grouped sample standard
deviations ($\D_2$, dotted line), skewness and kurtosis ($\D_4$,
dashed line) to perform density estimation. An important improvement
is observed with the addition of the standard deviations in the
dataset. This is confirmed numerically by inspecting the evolution of
the root integrated mean squared error (RIMSE) and of the
Kullback-Leibler (K-L) divergence between the 'target' density and the
estimate obtained using sample moments of increasing orders, see Table
\ref{PaymentData:FitStats:Tab}. When all the tabulated sample moments
are used (with $\D_4$), one can see (from the dashed curve in the
bottom of Fig.\,\ref{PaymentData:Fit:Fig} and from the K-L divergence
in Table \ref{PaymentData:FitStats:Tab}) that the target density is
nearly perfectly reconstructed.  The table also provides information
on the effective number of spline parameters $\mathrm{edf}(\hat\lambda)$
for the roughness penalty parameter $\hat\lambda$ selected using
Algorithm \ref{EM_FreqCentralMoments:Algo} and quantifying
the complexity of the density estimate. The fitted central moments can
be compared to their observed counterparts within each of the 3
classes, see Table \ref{PaymentData:FittedMoments:Tab}.  The observed
differences are within the sampling tolerances tuned by the
variance-covariance matrix $\Sigma_j$ of the moment estimators, see
(\ref{Sigma_j:Eq}), with a larger tolerance for class $\mathcal{C}_3$
as it is associated to the smallest frequency, $n_3=116$.
Value-at-Risk measures corresponding to 95\% and 99\% quantile
estimates were also computed using the theory of Section
\ref{QuantileEstimation:Sec} with 95\% credible intervals first
evaluated on the $\log_{10}$-scale using
(\ref{QuantileCredibleInterval:Eq}) and transformed back to the
original scale (in euros) for reporting purposes.  These can be
compared with the actual values that were calculated from the
confidential raw data, $\VaR_{5\%}=16\,125$ and $\VaR_{1\%}=38\,099$
euros, respectively.

\begin{table}[ht]
\centering
\begin{tabular}{ccccccccc}
  \toprule
          &            &               &             & \multicolumn{2}{c}{$\VaR_{5\%}$} && \multicolumn{2}{c}{$\VaR_{1\%}$} \\
  \cline{5-6} \cline{8-9}
Data $\D$ & \mcc{edf}  &  \mcc{RIMSE}  &  \mcc{K-L}  & Est. & 95\% cred.int. && Est. & 95\% cred.int.\\
\midrule
$\D_0$ &  6.2 & 0.069 & 0.042 & 16\,250 & (14\,795,\,17\,848) && 34\,764 & (29\,724,\,40\,658) \\
$\D_1$ &  6.7 & 0.030 & 0.029 & 15\,885 & (14\,617,\,17\,263) && 41\,502 & (37\,064,\,46\,472) \\
$\D_2$ &  9.0 & 0.027 & 0.019 & 16\,641 & (15\,355,\,17\,647) && 40\,766 & (35\,261,\,47\,131) \\
$\D_4$ & 11.7 & 0.012 & 0.001 & 16\,106 & (14\,896,\,17\,413) && 38\,988 & (33\,504,\,45\,371)\\
\bottomrule
\end{tabular}
\caption{Car insurance data: numerical comparison of density and Value-at-Risk
  estimates obtained using tabulated sample moments of increasing orders.}
\label{PaymentData:FitStats:Tab}
\end{table}

\begin{table}[ht]
\centering
\begin{tabular}{crrrrrrrrrrrrrr}
\toprule
  && \multicolumn{12}{c}{Central moments for $\log_{10}$(Claim)} \\
    \cline{4-14}
  Interval      & Freq. && \multicolumn{2}{c}{$M_1$} && \multicolumn{2}{c}{$M_2$} && \multicolumn{2}{c}{$M_3$} && \multicolumn{2}{c}{$M_4$}\\
\cline{4-5} \cline{7-8} \cline{10-11} \cline{13-14}
$\mathcal{C}_j$ & $n_j$ && Obs.  &Fitted && Obs.  &Fitted && Obs.  &Fitted && Obs.  & Fitted \\
\midrule                  
(0.00\,;\,3.00] & 1168 && 2.462 & 2.472 && 0.336 & 0.336 &&-0.350 &-0.351 && 0.611 & 0.619 \\
(3.00\,;\,4.30] & 2234 && 3.529 & 3.532 && 0.113 & 0.111 && 0.014 & 0.013 && 0.028 & 0.026 \\
(4.30\,;\,6.18] &  116 && 4.556 & 4.549 && 0.075 & 0.073 && 0.054 & 0.051 && 0.071 & 0.064 \\
\bottomrule
\end{tabular}
\caption{Car insurance data: observed and fitted central moments within classes using
  a model based on $\D_4$.}
\label{PaymentData:FittedMoments:Tab}
\end{table}

\section{Discussion}
We have shown how to combine tabulated summary statistics involving
moments of order one to four with the observed frequencies to estimate
a density from grouped data. The proposed inference strategy,
implemented in the R-package \texttt{degross}, relies on
an EM algorithm with uncertainty measures computed in a final step
from the observed penalized log-likelihood.  The penalty not only
encourages smoothing of the resulting density estimate, but also
ensures agreement up to sampling errors between the underlying
theoretical moments and their observed values in each class.  Simple
parametric alternatives might be considered for the density model in
specific settings. The nonparametric estimation studied here could
then be used to validate or select such proposals, or to point out
their possible shortcomings.

Although the transmission of data using tabulated summary statistics
may not be fully compliant with the European General Data Protection
Regulation (GDPR, EU 2016/679) guidelines, it enables to mask data
details by summarizing them with a couple of technical numbers besides
the class frequencies, see e.g.\,Table \ref{Payment:Tab}.  This is a
convenient method to communicate in a fairly accurate and compact way
on the distribution of the underlying raw data with a limited loss of
information.

That methodology might be combined with regression models where
information on the distribution of the response is provided in such a
summarized way conditionnally on a selected and limited number of
subject characteristics (such as the age category in the car insurance
example). At the individual level, besides covariates values, the
reported loss would take the form of a class indicator. The
challenge would be to make inference on the regression model
components from such imprecise information on the response data. Flexible
forms for the error distribution and for the quantification of
covariate effect on the response conditional distribution should be
compatible with the available information at the aggregate level.

\section*{Acknowledgments}

The author would like to thank Dr.\,Bernard Lejeune (ULiege, Belgium)
for useful discussions about the Generalized Method of Moments and
Prof.\,Michel Denuit (UCLouvain, Belgium) for motivating this project
and sharing the data used in the application.
Philippe Lambert also acknowledges the
support of the ARC project IMAL (grant 20/25-107) financed by the
Wallonia-Brussels Federation and granted by the Acad\'emie
Universitaire Louvain.

% \bibliographystyle{Chicago}
% \bibliography{SmoothHistogram}

\newpage
\printbibliography

\end{document}